\documentclass[12pt]{article}
\usepackage{amsmath,amsfonts}
\usepackage{graphicx}
\usepackage{booktabs}
\usepackage[left = 1.5cm, right = 1.5cm, top = 1.5cm, bottom = 1.5cm, footskip = 1cm]{geometry}
\usepackage{multirow}
\usepackage{authblk}
\usepackage{subcaption}

\usepackage[backend = biber, style = apa]{biblatex}
\title{Modelling and Forecasting of Complex Multiple Seasonalities in Categorical Time Series using TSOLR and ISOLR Models}

\author[]{\small{Anirban Ghosh}}
\author[]{Raju Maiti\thanks{Corresponding author: Raju Maiti, Indian Statistical Institute, 203 B.T. Road, Kolkata 700108, India. Email: raju.isical@gmail.com}}

\affil[]{Indian Statistical Institute, 203 B.T. Road, Kolkata 700108, India.}

\date{}

\begin{document}

\maketitle

\begin{abstract}
Multiple seasonalities have been widely studied in continuous time series using models such as TBATS, for instance in electricity demand forecasting. However, their treatment in categorical time series, such as air quality index (AQI) data, remains limited. Categorical AQI often exhibits distinct seasonal patterns at multiple frequencies, which are not captured by standard models. In this paper, we propose a framework that models multiple seasonalities using Fourier series and indicator functions, inspired by the TBATS methodology. The approach accommodates the ordinal nature of AQI categories while explicitly capturing weekly, monthly and yearly seasonal cycles. Simulation studies demonstrate the empirical consistency of parameter estimates under the proposed model. We further illustrate its applicability using real categorical AQI data from Kolkata and compare forecasting performance with Markov models and machine learning methods. Results indicate that our approach effectively captures complex seasonal dynamics and provides improved predictive accuracy. The proposed methodology offers a flexible and interpretable framework for analyzing categorical time series exhibiting multiple seasonal patterns, with potential applications in air quality monitoring, energy consumption and other environmental domains.
\end{abstract}

\noindent\textbf{Keywords:} \textit{Air Quality Index; Forecasting; Categorical Time Series; Multiple Seasonality; Fourier series; TBATS model.}

\section{Introduction}
Seasonality refers to the presence of systematic and repeating patterns in a time series at fixed intervals. In the context of continuous-valued series, a single seasonality arises when there is one dominant repeating cycle, such as the daily pattern in hourly electricity demand or the annual cycle in temperature data. More complex time series, however, often exhibit multiple seasonalities, where different seasonal cycles coexist at different frequencies. A prominent example is electricity consumption, which may display daily, weekly and annual seasonal patterns simultaneously. Models such as TBATS (\cite{Livera2011}) and its variants (\cite{Wang2025, Makatjane2024, Thayyib2023}) have been specifically developed to capture such multi-seasonal structures in continuous time series.

When the time series is categorical, such as daily Air Quality Index (AQI) data categorized into Good, Satisfactory, Moderate, Poor, Very Poor and Severe, the definition and treatment of seasonality becomes less straightforward. In this context, seasonality may be understood as systematic variation in the probability distribution of categories across temporal cycles. For example, in many Indian cities, the probability of observing Unhealthy AQI levels is higher during the winter months compared to the monsoon season. Similarly, at finer temporal resolutions, one might observe daily or weekly patterns in the distribution of categories. Extending this reasoning, multiple seasonalities in categorical time series refer to the coexistence of such repeating patterns across more than one time scale, such as daily, weekly and annual cycles in AQI categories.

Modeling seasonality in categorical time series presents unique challenges, as standard approaches designed for continuous data cannot be directly applied. Existing methods include Markov chain models that capture transition probabilities between categories, generalized linear models for categorical responses (e.g., logistic regression with seasonal covariates) and more recently, machine learning methods such as random forests or recurrent neural networks. However, these methods often fail to explicitly account for multiple seasonal cycles. Incorporating flexible representations such as Fourier series expansions or indicator functions into categorical models provides a promising direction for capturing multi-seasonal structures, inspired by the success of models like TBATS in the continuous domain.

To address these challenges, we propose an extension of the TBATS framework tailored for categorical time series. Specifically, we employ a proportional odds ordinal logistic regression model (\cite{Agresti2002}) to account for the ordered nature of AQI categories, while incorporating Fourier series terms as covariates to flexibly represent multiple seasonal patterns at daily, weekly and annual frequencies. This design preserves the interpretability of seasonal effects, as the Fourier coefficients directly reflect the contribution of each cycle, while ensuring consistency with the ordinal data structure. 

As an alternative to Fourier-based representations, multiple seasonalities can also be modeled using indicator functions that directly encode recurring time segments. For example, separate dummy variables may be introduced for each hour of the day, day of the week, or month of the year, thereby capturing seasonal effects in a fully nonparametric manner. Unlike Fourier series, which approximate seasonal cycles smoothly using sinusoidal functions, indicator functions allow for sharp changes and irregular seasonal shapes that are often observed in practice, such as sudden spikes in AQI levels during specific festival days or winter months. This approach provides greater flexibility in representing heterogeneous seasonal effects across categories, though it typically requires a larger number of parameters and may risk overfitting unless carefully regularized. Nevertheless, indicator-based modeling offers a complementary perspective to Fourier expansions and can be especially useful when the seasonal structure is highly irregular or difficult to approximate with smooth harmonic terms.

The remainder of this article is structured as follows. Section \ref{sec:background} provides a comprehensive review of existing research on AQI modeling and forecasting techniques. Section \ref{sec:data_analysis} describes the dataset and includes an exploratory analysis that motivates the modeling approach. The proposed ordinal logistic regression models are introduced and detailed in Section \ref{sec:logistic}. Section \ref{sec:mle} details the estimation procedure of the proposed models. Some existing methods for modeling and forecasting categorical time series data are summarized in Section \ref{sec:methods}. Section \ref{sec:autoassociation} discusses the procedure for selecting order of the model based on various auto-association measures. Simulation experiments assessing the estimation accuracy and forecasting capabilities of the proposed models are reported in Section \ref{sec:simulation}. Section \ref{sec:results} presents AQI data analysis results. Finally, Section \ref{sec:discussion} concludes the article and suggests directions for future research.

\section{Literature Review}      \label{sec:background}
Numerous studies have focused on modeling air quality using time series techniques, with most treating AQI as a continuous variable and applying models such as ARIMA, VAR and state-space approaches (\cite{Sethi2020, Liu2021, Gupta2023b, Liu2024}). While these methods are effective for continuous data, AQI is commonly reported in ordinal categories for public health communication, making continuous-value models less directly interpretable. Research on categorical time series models remains relatively limited (\cite{Liu2022, weiss2025}). Classical Markov chain models have been widely used to capture transitions between AQI categories, but they suffer from exponential growth in the number of parameters as the state space increases. To mitigate this, extensions such as discrete ARMA (DARMA) processes (\cite{Jacobs1978a, Jacobs1978b, Jacobs1978c}), mixture transition distribution (MTD) models (\cite{Raftery1985, Berchtold2002}), Pegram’s operator-based ARMA process (\cite{Pegram1980, biswas2009song}) and regime-switching DARMA models (\cite{weiss2020}) have been proposed. However, these approaches largely fail to address multiple seasonalities, which are a defining feature of urban AQI series, nor do they flexibly incorporate exogenous drivers such as meteorological variables.

In parallel, machine learning methods, like random forests, support vector machines and deep learning, have shown strong predictive performance in air quality forecasting (e.g., \cite{Gupta2023a, Mishra2024, Desai2025}). Yet, these models are often criticized for limited interpretability and a lack of explicit representation of seasonal structures. By contrast, ordinal logistic regression (\cite{Agresti2002, Fokianos2003}) offers a statistically grounded and interpretable framework for modeling ordered categorical outcomes, while naturally accommodating external covariates such as temperature, humidity, traffic intensity and industrial activity. Its flexibility also allows for the incorporation of seasonal components, for instance through Fourier expansions or indicator functions. Despite this potential, the use of ordinal regression models to explicitly model multiple seasonalities in categorical AQI series remains underexplored, particularly in urban contexts like Kolkata where both anthropogenic factors and climatic cycles strongly influence air quality dynamics.

\section{Daily AQI Data of Kolkata}     \label{sec:data_analysis}
The air quality index (AQI) is an indicator for reporting air quality daily. In India, the AQI is calculated using the concentrations of 12 types of air pollutants ascertained by Central Pollution Control Board (CPCB). For each pollutant, a sub-index,defined by CPCB, is calculated using its measured concentration. The overall AQI is the maximum of the individual sub-indices across all available pollutants. Detailed description of calculating AQI in India is discussed in \cite{WBPCB2022}.

\subsection{Study Area}
\begin{figure}
    \centering
    \includegraphics{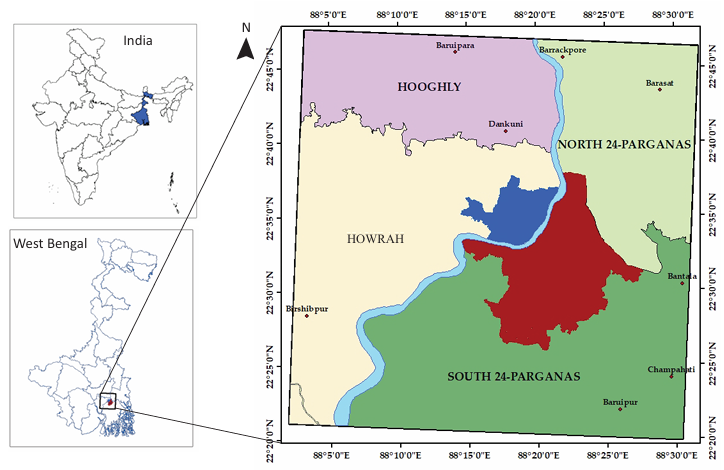}
    \caption{Location map of Kolkata city (red-shaded region) in India (\cite{Dutta2025map})}
    \label{fig:map}
\end{figure}

This study focuses on the air quality of Kolkata, the capital city of the Indian state of West Bengal. Kolkata is the third most populous metropolitan region in India, following Delhi and Mumbai, encompassing a total area of 206.08~km$^{2}$ with a population of approximately 4.5 million and a population density of about 30,000~persons per km$^{2}$ (\url{https://en.wikipedia.org/wiki/Kolkata}). Geographically, the city is situated at 22$^{\circ}$34$^{\prime}$3$^{\prime\prime}$N and 88$^{\circ}$22$^{\prime}$12$^{\prime\prime}$E. The geographical position of Kolkata within India is illustrated in Figure~\ref{fig:map}. It serves as one of the major industrial and economic hubs of the country, contributing significantly to regional development. Among the twelve nationally monitored pollutants, only particulate matters $PM_{10}$ and $PM_{2.5}$ have been observed to exceed permissible limits during the winter season in Kolkata (see, e.g., \cite{WBPCB2022}).

\subsection{AQI Dataset}
For this study, daily categorical AQI data for Kolkata is considered over a multiyear period, available from 2019 onward. The data of 2019-2024 is collected from official air quality monitoring stations operated by the Central Pollution Control Board (CPCB) and the West Bengal Pollution Control Board (WBPCB). These stations record daily pollutant concentrations. Data are publicly available through the AQI Data Repository of National Air Quality Index (NAQI) portal (\url{https://airquality.cpcb.gov.in/ccr/#/caaqm-dashboard-all/caaqm-landing/aqi-repository}), which provides a reliable basis for time series analysis and forecasting.

\subsection{Data Preprocessing and Preparation}
For model fitting, the raw AQI data are first preprocessed to address missing or inconsistent entries. The final daily AQI values are then mapped to standard categorical labels following the National AQI classification scheme (\url{https://www.iitk.ac.in/new/national-air-quality-index}), as shown in Table \ref{tab:category}. These categorical labels are subsequently encoded numerically in ordinal form to facilitate statistical modeling. The clean and ready-to-use daily AQI data of Kolkata from 2019 to 2024, along with the associated R and Python codes for data analysis, are available on the author's GitHub repository: \url{https://github.com/ganirban004/aqi-logistic-paper.git}.

\begin{table}[htpb]
    \centering
    \caption{National AQI Classification Scheme}
    \label{tab:category}
    \begin{tabular}{|l|l|c|}
    \hline 
    \textbf{Avearge AQI value over 24 hours } & \textbf{Air Quality Category} & \textbf{Code} \\ \hline 
    $\leq$ 50                                 & Good                          & 0 \\ 
    51-100                                    & Satisfactory                  & 1 \\ 
    101-200                                   & Moderate                      & 2 \\ 
    201-300                                   & Poor                          & 3 \\
    301-400                                   & Very Poor                     & 4 \\
    $\ge$ 401                                 & Severe                        & 5 \\ \hline
    \end{tabular} 
\end{table}

\subsection{Exploratory Analysis}
\begin{figure}[htpb]
    \centering
    \begin{subfigure}[htpb]{0.48\textwidth}
        \centering
        \includegraphics[scale = 0.96]{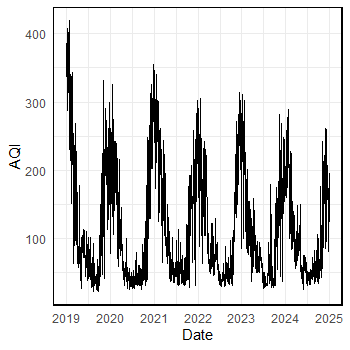}
        \caption{Daily Air Quality Index (AQI) time series of Kolkata during 2019-2024 (higher values indicate poorer air quality)}
        \label{fig:original_AQI}
    \end{subfigure}
    \hfill
    \begin{subfigure}[htpb]{0.51\textwidth}
        \centering
        \includegraphics[scale = 0.96]{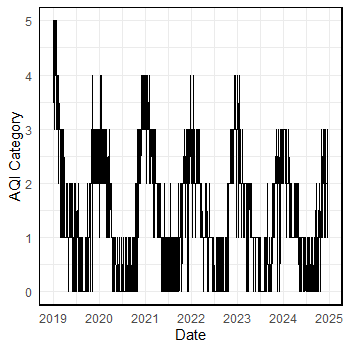}
        \caption{Daily categorical AQI time series for Kolkata from 2019 to 2024}
        \label{fig:categorical_AQI}
    \end{subfigure}
    \caption{Daily Air Quality Index (AQI) time series (original and categorical) of Kolkata during 2019-2024 showing the variation in air pollution levels over time with noticeable seasonal and episodic fluctuations}
    \label{fig:full_aqi}
\end{figure}

Exploratory data analysis is performed to understand the distribution and temporal patterns of AQI categories in Kolkata. The original numerical AQI time series and the corresponding transformed categorical time series are presented in Figures \ref{fig:original_AQI} and \ref{fig:categorical_AQI}, respectively. Both representations exhibit a distinct seasonal pattern in air quality across the year. A clear decreasing trend is observed in the numerical AQI values, indicating gradual improvement in air quality. In the categorical series (Figure \ref{fig:categorical_AQI}), no instance of the ``Severe'' category is observed and notably, the last two winters show an absence of ``Very Poor'' days as well. These observations collectively suggest a substantial improvement in Kolkata’s air quality in recent years.

\begin{figure}[htpb]
    \centering
    \begin{subfigure}[htpb]{0.48\textwidth}
        \centering
        \includegraphics[scale = 0.7]{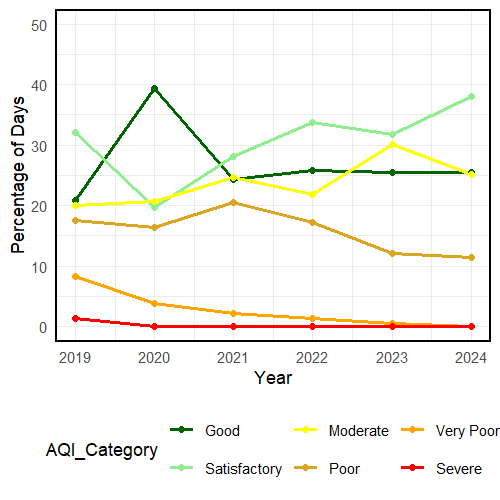}
        \caption{Yearly Trend of categorical AQI levels in Kolkata from 2019 to 2024}
    \label{fig:yearly}
    \end{subfigure}
    \hfill
    \begin{subfigure}[htpb]{0.48\textwidth}
        \centering
        \includegraphics[scale = 0.7]{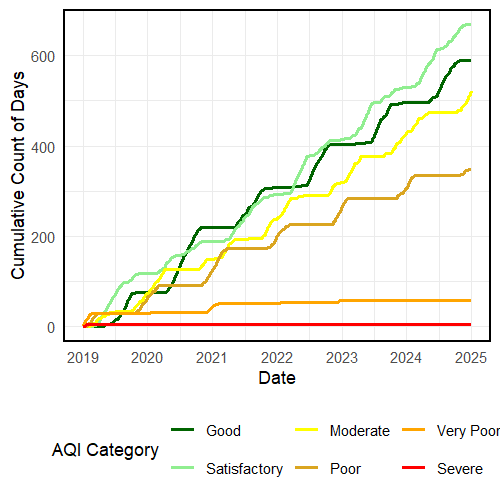}
        \caption{Daily rate evolution graph of categorical AQI levels in Kolkata from 2019-2024}
    \label{fig:rateevolution}
    \end{subfigure}
    \caption{Yearly trend and rate evolution graph of categorical AQI levels in Kolkata from 2019-2024}
\end{figure}

The temporal evolution of air quality in Kolkata is depicted in Figure \ref{fig:yearly}, showing the annual percentage distribution of days across different AQI categories from 2019 to 2024. A clear upward trend is observed in the proportion of ``Satisfactory'' days, indicating progressive improvement in air quality over the study period, except for the year 2020. During 2020, corresponding to the COVID-19 lockdown, there is a sharp increase in the proportion of ``Good'' AQI days, attributable to reduced industrial and vehicular activity. In contrast, the shares of ``Poor'' and ``Very Poor'' categories show a steady decline and notably, no ``Severe'' AQI days are recorded after 2019 as seen in Figure \ref{fig:categorical_AQI}. Figure \ref{fig:rateevolution} shows the evolution of rate of different AQI categories over time (see \cite{weiss2020} to know more about rate evolution graphs). Slope of this rate evolution graph gives the estimated marginal probability of the corresponding category. The curve for ``Very Poor" and ``Severe" categories are almost flattened in recent years. Overall, these findings also reflect a significant and consistent enhancement in Kolkata’s air quality in recent years in support of Figure \ref{fig:full_aqi}.

\begin{figure}[htpb]
    \centering
    \begin{subfigure}[htpb]{0.40\textwidth}
        \centering
        \includegraphics[scale = 0.9]{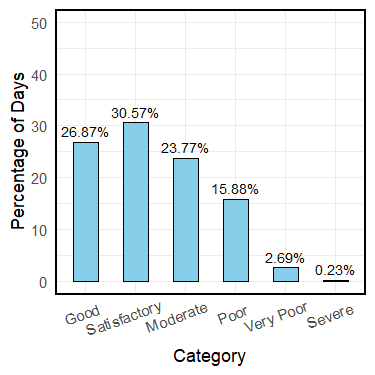}
        \caption{Frequency distribution of AQI categories in Kolkata over the entire study period (2019–2024)}
        \label{fig:frequency}
    \end{subfigure}
    \hfill
    \begin{subfigure}[htpb]{0.58\textwidth}
        \centering
        \caption{Transition Probability Matrix of Air Quality Categories over the entire study period (2019-2024)}
        \label{tab:transition-matrix}
        \begin{tabular}{|l|cccccc|}
            \hline
            \textbf{Cat.} & \textbf{Good} & \textbf{Sat.} & \textbf{Mod.} & \textbf{Poor} & \textbf{V.P.} & \textbf{Sev.} \\
            \hline
            Good         & 0.80 & 0.19 & 0.00 & 0.00 & 0.00 & 0.00 \\
            Sat.         & 0.17 & 0.70 & 0.13 & 0.00 & 0.00 & 0.00 \\
            Mod.         & 0.01 & 0.16 & 0.68 & 0.15 & 0.00 & 0.00 \\
            Poor         & 0.00 & 0.01 & 0.22 & 0.70 & 0.07 & 0.00 \\
            V.P.         & 0.00 & 0.00 & 0.00 & 0.41 & 0.53 & 0.07 \\
            Sev.         & 0.00 & 0.00 & 0.00 & 0.00 & 0.80 & 0.20 \\
            \hline
        \end{tabular}
    \end{subfigure}
    \caption{Frequency distribution of AQI categories and Transition probability matrix for the categorical AQI time series.}
\end{figure}

The frequency plot shown in Figure \ref{fig:frequency} illustrates the prevalence of each air quality level throughout the study period (2019-2024). The transition probability matrix given in Table \ref{tab:transition-matrix} shows the probabilities of transitioning from one air quality category to another on consecutive days during the study period. Higher values along the diagonal indicate strong temporal persistence within the same category, while non-zero off-diagonal entries reflect the likelihood of short-term changes in air quality conditions.

\subsection{Month-Category Heatmap using Intensity Measure}
To visualize the seasonal patterns of categorical AQI data, we construct a month-category heatmap using an intensity measure. Let $Y_{t} \in \{0, 1, 2, 3, 4, 5\}$ denote the AQI category on day $t$, encoded as Good = 0, Satisfactory = 1, Moderate = 2, Poor = 3, Very Poor = 4 and Severe = 5. Let $m \in \{1, \cdots, 12\}$ denote the month of observation. For each month $m$ and AQI category $j$, we define the frequency of occurrence as
\[
f_{m,c} = \sum_{t \in \text{Month } m} \mathbb{I}(Y_{t} = j),
\]
where $\mathbb{I}(\cdot)$ is the indicator function. Alternatively, one may use the percentage of days in each category by dividing $f_{m,c}$ by the total number of days in month $m$ and then multiply by 100.  

\begin{figure}
    \centering
    \includegraphics[scale = 0.9]{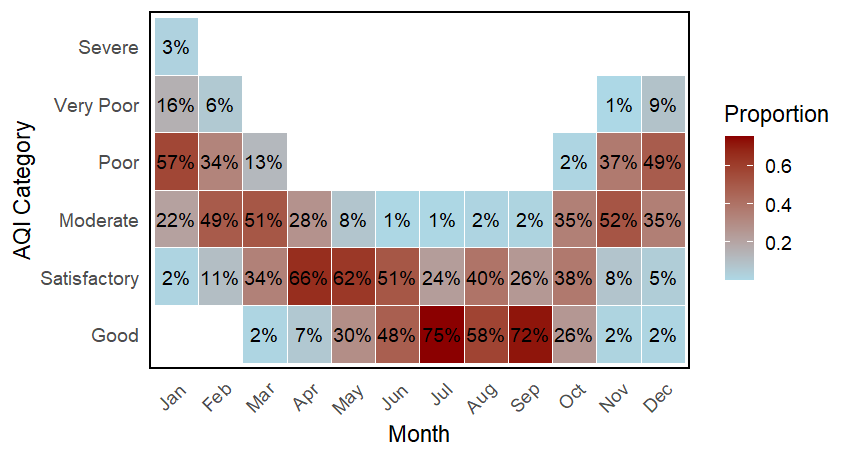}
    \caption{Month-category heatmap of AQI categories using intensity measure}
    \label{fig:heatmap}
\end{figure}

The heatmap is constructed by placing months along the horizontal axis and AQI categories along the vertical axis, with each cell’s color intensity representing the corresponding percentage of days. Figure \ref{fig:heatmap} illustrates this visualization. Darker shades denote higher percentages, offering an immediate visual impression of the distribution of air quality levels across different months. This graphical representation effectively highlights seasonal patterns, such as the increased frequency of Poor and Very Poor air quality days during winter and the predominance of Moderate or Good air quality days throughout the Monsoon season.

\subsection{Multiple Seasonality of AQI Data}

Here, we explored the seasonality of the data at different frequencies using category proportions, Month-Category heatmaps and recovery time analysis.

\subsubsection{Seasonal Patterns: Summer vs Monsoon vs Autumn vs Winter}
\begin{figure}[htpb]
    \centering
    \includegraphics[scale = 0.9]{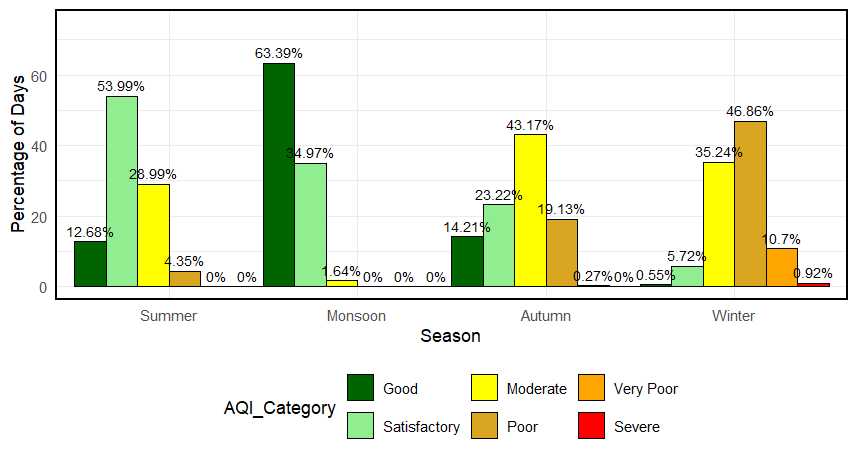}
    \caption{Seasonal distribution of categorical AQI levels in Kolkata from 2019 to 2024}
    \label{fig:seasonal}
\end{figure}

The plot, displayed in Figure \ref{fig:seasonal}, illustrates the variation in air quality across the four seasons namely summer, monsoon, autumn and winter. Poorer air quality levels (e.g., ``Moderate", ``Poor" and ``Very Poor") are more prevalent during winter and post-monsoon seasons, while cleaner air conditions (e.g., ``Good" and ``Satisfactory") are more common during the monsoon, highlighting the seasonal influence on pollution dynamics. The monsoon season typically exhibits better air quality, likely due to the cleansing effect of rainfall. These recurring seasonal fluctuations suggest the need to incorporate seasonal components or dummy variables when modeling time series.

\subsubsection{Seasonal Patterns: Weekdays vs Weekends}
\begin{figure}
    \centering
    \includegraphics[scale = 0.9]{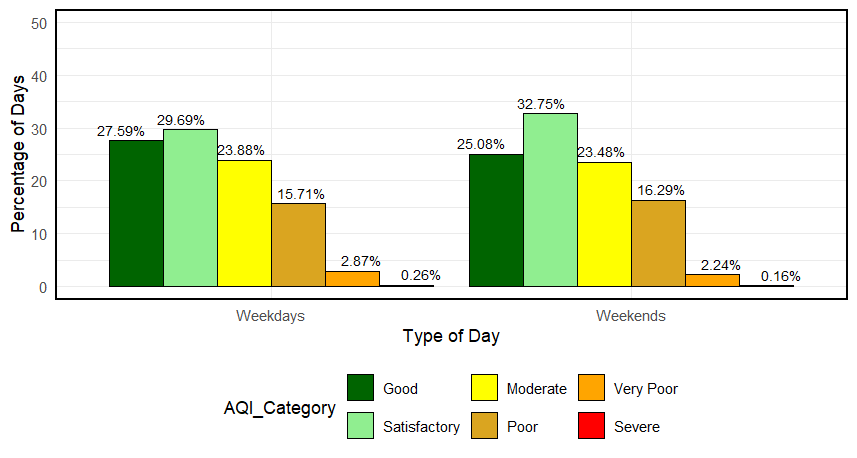}
    \caption{Frequency distribution of AQI categories during weekdays and weekends in Kolkata (2019-2024)}
    \label{fig:daytype}
\end{figure}

The distribution of AQI categories across weekdays and weekends, depicted in Figure \ref{fig:daytype}, exhibits subtle yet meaningful variations. On weekends, a higher proportion of days falls under the ``Satisfactory'' category (32.75\%) compared to weekdays (29.69\%), suggesting slightly better air conditions during weekends. Meanwhile, the proportion of ``Good'' days is marginally higher on weekdays (27.59\%) than weekends (25.08\%). The frequencies of ``Moderate'' and ``Poor'' categories remain relatively consistent across both types of day, while the most hazardous categories, like ``Very Poor'' and ``Severe'', occurred infrequently, with slightly higher occurrences on weekdays. These findings indicate that although air quality remains broadly similar throughout the week, weekends show a modest shift toward improved conditions, possibly due to decreased traffic and industrial activity.

\subsubsection{Seasonal Patterns: Festive vs Non-festive}
Air pollution in Kolkata displays pronounced seasonal variability, with significant intensification observed during the Diwali festival period. In this study, we examine the specific impact of Diwali on air quality in the city. To analyze this, the study period is divided into three sequential phases as follows:
\begin{itemize}
    \item \textbf{Pre-Diwali Period (Day $-$22 to Day $-$8):}  
    This period refers to the 15 days leading up to the Diwali window, specifically from 22 days before Diwali to 8 days before Diwali.  
    \[
    \text{Pre-Diwali Period} = [\text{Diwali} - 22\ \text{days},\ \text{Diwali} - 8\ \text{days}]
    \]
    \item \textbf{Diwali Period (Day $-$7 to Day $+$7):}  
    This is a 15-day symmetric window centered on the Diwali festival, including 7 days before and 7 days after Diwali.  
    \[
    \text{Diwali Period} = [\text{Diwali} - 7\ \text{days},\ \text{Diwali} + 7\ \text{days}]
    \]
    \item \textbf{Post-Diwali Period (Day +8 to Day +22):}  
    This period includes the 15 days following the Diwali window, from 8 days after Diwali to 22 days after Diwali.  
    \[
    \text{Post-Diwali Period} = [\text{Diwali} + 8\ \text{days},\ \text{Diwali} + 22\ \text{days}]
    \]
\end{itemize}

\begin{figure}[htpb]
    \centering
    \includegraphics[scale = 0.9]{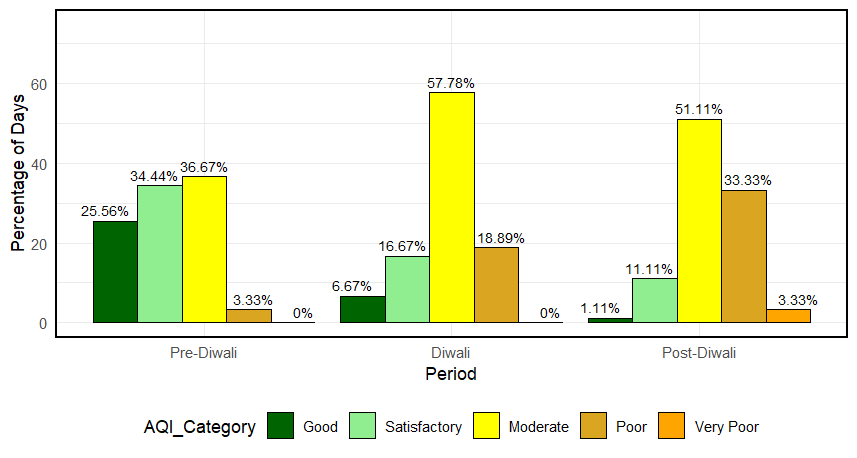}
    \caption{Frequency distribution of AQI categories during the Pre-Diwali, Diwali and Post-Diwali periods in Kolkata (2019-2024)}
    \label{fig:festivity}
\end{figure}

This sequential structure allows for systematic analysis of air quality trends before, during and after the Diwali festival, isolating the impact of firecracker emissions and seasonal variations. The plot displayed in Figure \ref{fig:festivity} highlights a noticeable increase in the frequency of ``Moderate", ``Poor" and ``Very Poor” air quality days during the Diwali and Post-Diwali periods, indicating a significant short-term deterioration in air quality associated with festival-related emissions.

In summary, the categorical time-series analysis offers clear evidence of multiple forms of seasonality in Kolkata’s AQI categories. Properly modelling these seasonal components is therefore essential for improving the accuracy of pollution assessment and supporting effective public-health and mitigation strategies.

\section{Proposed Model}  \label{sec:logistic}
In the context of electricity consumption data in Australia, \cite{Livera2011}, proposed the TBATS model to capture multiple seasonalities and trend. The TBATS model is designed to handle complex time series with multiple seasonalities, nonlinear transformations and short-term dependencies. The acronym TBATS stands for Trigonometric seasonality, Box--Cox transformation, ARMA errors, Trend and Seasonal components. Let $y_t$ be the observed value of the time series at time point $t$. A Box-Cox transformation is first applied on the original time series $y_t$ as
\begin{equation*}
    y_{t}^{(\omega)} = 
    \begin{cases}
        \dfrac{y_{t}^\omega - 1}{\omega} & \text{for } \omega \neq 0 \\ 
        \log y_{t} & \text{for } \omega = 0
    \end{cases}
\end{equation*}
where $\omega$ is the transformation parameter. The transformed series is then modeled as
\begin{equation*}
    y_{t}^{(\omega)} = l_{t} + \phi b_{t} + \sum_{i=1}^{T} s_{i, t - P_{i}} + \varepsilon_{t},
\end{equation*}
where $l_{t}$ is the local level, $b_{t}$ is the trend component with damping parameter $\phi \in (0, 1)$, $s_{i, t}$ are the seasonal components, $T$ is the number of seasonal components and $\varepsilon_{t}$ follows an ARMA($p,q$) process. Each seasonal component is expressed using trigonometric terms based on a Fourier expansion. For a seasonal period $P_{i}$, the representation is
\begin{equation*}
    s_{i, t} = \sum_{k=1}^{K_i} \left[ \beta_{1i}^{(k)} \cos\left(\frac{2 \pi k t}{P_{i}}\right) + \beta_{2i}^{(k)} \sin\left(\frac{2 \pi k t}{P_{i}}\right) \right]
\end{equation*}
where $K_i$ is the number of harmonics chosen for seasonal cycle $i$. The state vectors $\beta_{1i}^{(k)}$ and $\beta_{2i}^{(k)}$ evolve dynamically according to recursive trigonometric updates. In its original formulation, the TBATS model effectively captures multiple seasonalities, nonlinear trends and short-term dependencies in continuous-valued time series, such as electricity demand in Australia (\cite{Livera2011}). The use of Fourier terms for trigonometric seasonality, Box–Cox transformations and ARMA error structures allows TBATS to model complex periodic patterns across multiple timescales simultaneously. However, its direct applicability to categorical time series is limited because TBATS assumes continuous observations and additive errors, whereas categorical AQI data are discrete and ordinal.

To bridge this gap, we propose an extension of the TBATS idea for ordinal time series. Specifically, we employ a proportional odds model (\cite{Agresti2002}) for ordinal responses, with Fourier series terms as covariates to capture multiple seasonal patterns hidden in the AQI data. Accordingly, we refer to this framework as the Trigonometric Seasonality in Ordinal Logistic Regression (TSOLR) model. This approach preserves the key strength of TBATS, explicit modeling of multiple seasonal cycles, while respecting the discrete, ordered nature of AQI categories.

Let $Y_{t} \in \{0, 1, \cdots, m\}$ denote the AQI category on day $t$. We assume that the underlying latent process is influenced by multiple seasonalities, analogous to the TBATS framework, but adapted to an ordinal regression setting.  We adopt the proportional odds model (\cite{Agresti2002}), where the cumulative probability of being in category $j$ or below at time $t$ is given by
\begin{equation}    \label{equ:logistic_general}
    \text{logit} \Big( P(Y_t \leq j \mid \mathcal{F}_{t - 1}, \boldsymbol{X_{t}}) \Big) = \theta_{j} - \eta_{t}, \quad j = 0, 1, \cdots, m - 1
\end{equation}
where $\mathcal{F}_{t}$ contains all the past information up to time point $t$, i.e., $\mathcal{F}_{t} = \{(y_{t^{'}}, \boldsymbol{X_{t^{'}}}), t^{'} =1, 2, \cdots, t\}$, $\theta_{j}$ are category-specific threshold parameters and $\eta_{t}$ is the linear predictor. In particular, to capture multiple seasonalities using Fourier terms as covariates, $\eta_{t}$ takes the form
\begin{equation*}
    \eta_{t} = \sum_{i=1}^{T} \sum_{k=1}^{K_i} \Big[ \beta_{1i}^{(k)} \cos \Big( \frac{2 \pi k t}{P_i} \Big) + \beta_{2i}^{(k)} \sin \Big( \frac{2 \pi k t}{P_{i}}\Big) \Big] + \boldsymbol{X_{t}^{\top}\boldsymbol{\gamma}},
\end{equation*}
where $P_i$ is the seasonal period for $i^{th}$ seasonal component (e.g., in daily data, $P_{i}$ = 7 for weekly seasonality, $P_{i}$ = 365 for yearly seasonality etc.) and $K_i$ is the number of Fourier harmonics chosen for a particular seasonal cycle $i$. To avoid overfitting and to keep the model simple and interpretable, we assume that $K_{i}$ = 1 and it well captures the seasonality through a single sine-cosine curve. $\beta_{1i}$ and $\beta_{2i}$ (as $K_{i}$ = 1, the superscript $(k)$ is omitted) are Fourier coefficients corresponding to $i^{th}$ seasonal component. The term $\boldsymbol{X_{t}}$ includes additional exogenous covariates and $\boldsymbol{\gamma}$ be the corresponding parameter vector. Then the TSOLR model takes the form
\begin{equation}    \label{equ:logistic_fourier}
    \text{logit} \Big( P(Y_t \leq j \mid \mathcal{F}_{t - 1}, \boldsymbol{X_{t}}) \Big) = \theta_{j} - \sum_{i = 1}^{T} \Big[ \beta_{1i} \cos \Big( \frac{2 \pi t}{P_i} \Big) + \beta_{2i} \sin \Big( \frac{2 \pi t}{P_{i}}\Big) \Big] + \boldsymbol{X_{t}^{\top}} \boldsymbol{\gamma}
\end{equation}

This formulation allows the ordinal AQI model to inherit the TBATS philosophy of explicitly modeling multiple seasonal components, while remaining suitable for discrete, ordered outcomes. Short-term dependencies may further be incorporated by including lagged category indicators or autoregressive terms in $\eta_t$, analogous to ARMA errors in the TBATS model. Beside these, terms like $|\sin \left( \frac{2 \pi t}{P_{i}} \right)|$ and $|\cos \left( \frac{2 \pi t}{P_{i}} \right)|$ can also be included in the model to get non-negative seasonal effects.

In other classical approach, multiple seasonalities are captured through indicator variables which are discrete functions as compared to Fourier series. We refer to this framework as the Indicator Seasonality in Ordinal Logistic Regression (ISOLR) model. For the daily AQI data, there are three seasonalities: weekly, seasonal-seasonality and festival-seasonality (Diwali vs Non-Diwali). Hence, the linear predictor $\eta_{t}$ in model (\ref{equ:logistic_general}) takes the form
\begin{equation*}
    \eta_{t} = \beta_{1} \cdot Summer_{t} + \beta_{2} \cdot Monsoon_{t} + \beta_{3} \cdot Winter_{t} + \beta_{4} \cdot Diwali_{t} + \beta_{5} \cdot Weekend_{t} + \beta_{6} \cdot Y_{t - 1}
\end{equation*}
where $Summer_{t}$, $Monsoon_{t}$ and $Winter_{t}$ takes the value 1 if day $t$ falls in summer, monsoon and winter season respectively and 0 otherwise. The autumn season is considered as the baseline category. Similarly, $Diwali_{t}$ takes the value 1 if day $t$ falls in the 15-day Diwali period and $Weekend_{t}$ takes the value 1 if day $t$ is either Saturday or Sunday, 0 otherwise. $Y_{t - 1}$ is the lag-1 value at day $t$. Then the ISOLR model becomes
\begin{align} \label{equ:logistic_indicator}
    \log\left( \frac{P(Y_{t} \leq j)}{P(Y_{t} > j)} \right) 
    &= \theta_{j} 
    - \Big( \beta_{1} \cdot Summer_{t} + \beta_{2} \cdot Monsoon_{t} + \beta_{3} \cdot Winter_{t} \notag \\
    &\quad\;\;\, + \beta_{4} \cdot Diwali_{t} + \beta_{5} \cdot Weekend_{t} + \beta_{6} \cdot Y_{t - 1} \Big)
\end{align}

\section{Estimation of Parameters}     \label{sec:mle}
To estimate the parameters of the proposed model, standard practice is to maximize the log-likelihood function
\begin{equation*}
    \ell(\boldsymbol{\theta}, \boldsymbol{\beta}) = \log \prod_{t = 1}^{N} \prod_{j = 0}^{m} P(Y_{t} = j)^{\mathbb{I}(Y_{t} = j)} = \sum_{t = 1}^{N} \sum_{j = 0}^{m} \mathbb{I}(Y_{t} = j) \log P(Y_{t} = j)
\end{equation*}
where
\begin{equation*}
    P(Y_{t} = j) = 
    \begin{cases}
        P(Y_{t} \le j) & \text{for } j = 0 \\
        P(Y_{t} \le j) - P(Y_{t} \le j - 1) & \text{for } j = 1, 2, \cdots, m - 1 \\
        1 - P(Y_{t} \le m - 1) & \text{for } j = m
    \end{cases}
\end{equation*}
and
\begin{equation*}
    P(Y_{t} \le j) = 
    \begin{cases}
        \dfrac{\exp(\theta_{j} - \eta_t)}{1 + \sum_{i = 0}^{m - 1}\exp(\theta_{i} - \eta_t)} & \text{if } j = 0, 1, \cdots, m - 1 \\
        \dfrac{1}{1 + \sum_{i = 0}^{m - 1}\exp(\theta_{i} - \eta_t)} & \text{if } j = m
    \end{cases}
\end{equation*}
as described in model (\ref{equ:logistic_general}), with respect to $\boldsymbol{\theta}$ = ($\theta_{0}$, $\theta_{1}$, $\cdots$, $\theta_{m - 1}$) and $\boldsymbol{\beta}$. Under some regularity conditions, it is well established fact that the MLEs are asymptotically normal which is given by
\begin{equation*}
    \begin{bmatrix}
\boldsymbol{\hat{\theta}} \\
\boldsymbol{\hat{\beta}}
\end{bmatrix}
\overset{a}{\sim} 
\mathcal{N} \left(
\begin{bmatrix}
\boldsymbol{\theta} \\
\boldsymbol{\beta}
\end{bmatrix}, \;
\mathcal{I}^{-1}(\boldsymbol{\theta}, \boldsymbol{\beta})
\right).
\end{equation*}
The observed Fisher information matrix $\mathcal{I}(\theta)$ is computed as the negative of the Hessian matrix of the log-likelihood at the MLE:
\begin{equation*}
    \hat{\mathcal{I}}(\boldsymbol{\theta}, \boldsymbol{\beta}) = -\nabla^{2} \ell(\boldsymbol{\theta}, \boldsymbol{\beta}) \big|_{(\boldsymbol{\theta}, \boldsymbol{\beta}) = (\boldsymbol{\hat{\theta}}, \hat{\boldsymbol{\beta})}}
\end{equation*}
This Hessian matrix is computed numerically and it serves as the variance-covariance matrix of the estimated parameters, which can be further used to construct confidence interval of the estimated parameters. For parameter $(\boldsymbol{\theta}, \boldsymbol{\beta})$, the partial derivatives are approximated using finite differences of the log-likelihood function $\ell(\boldsymbol{\theta}, \boldsymbol{\beta})$. The step size for calculating finite differences is chosen to balance numerical precision and stability. The \texttt{polr()} function from \texttt{MASS} package in R implements the proportional odds cumulative logistic regression model.

\section{Existing Models}   \label{sec:methods}
In order to compare the forecasting performance of TSOLR and ISOLR models, we list some existing methodologies for modeling categorical time series in this section. The list includes some Markov chain based models along with some machine learning and deep learing models.
\subsection{Markov Chain Model}
A traditional approach for modeling categorical time series is the use of Markov Chains of order $p$, where the transition probability is defined as:
\begin{equation*}
    P(Y_{t} = j \mid Y_{t - 1}, \cdots, Y_{t - p}), \quad j = 0, 1, \cdots, m.
\end{equation*}

However, this approach suffers from an exponential increase in the number of parameters. For a Markov chain of order $p$ with $(m+1)$ categories, the number of parameters is $m(m+1)^p$. For example, with $m=4$ and $p=2$, 100 parameters are required, which is computationally intensive. Also this does not allow to incorporate external covariate in the model.

\subsection{Mixture Transition Distribution (MTD) Model}
To reduce complexity, the MTD model was introduced by \cite{Raftery1985} and later refined by \cite{Berchtold2002}. It models transition probabilities as a weighted mixture of one-step transitions:
\begin{equation*}
    P(Y_{t} \mid Y_{t - 1}, \cdots, Y_{t - p}) = \sum_{k = 1}^{p} \lambda_{k} P(Y_{t} \mid Y_{t - k}),
\end{equation*}
where $\lambda_{k} \geq 0$ and $\sum \lambda_{k} = 1$. The MTD model reduces the parameter count to $m(m+1) + p - 1$. For example, with $m=4$ and $p=2$, the parameter count drops to 21. Like the Markov chain model, this approach also does not consider information of any external covariate.

\subsection{Pegram's Autoregressive (PAR) Process}
\cite{Pegram1980} proposed a simplified approach called the PAR($p$) process, a special case of the discrete ARMA model (\cite{Jacobs1978a, Jacobs1978b, Jacobs1978c}). The PAR process is defined by:
\begin{equation*}
    P(Y_{t} = j \mid Y_{t - 1}, \cdots, Y_{t - p}) = \sum_{k = 1}^{p} \phi_{k} \mathbb{I}(Y_{t - k} = j) + \left(1 - \sum_{k = 1}^{p} \phi_{k} \right) \pi_{j},
\end{equation*}
where $\phi_{k} \in (0,1)$ and $\sum_{k = 1}^{p} \phi_{k} < 1$, $\pi_{j}$ is the marginal probability of category $j$ and $\mathbb{I}(\cdot)$ is the indicator function. However, this model does not include external covariate in the model.


\subsection{Decision Trees and Random Forests for Ordinal Data}
Standard tree-based methods can be adapted conveniently (\cite{Janitza2016}) using ordinal splitting criteria like the quadratic entropy index or the use of monotonic constraints. Random Forest (RF) is an ensemble-based learning method that extends traditional decision tree models by combining multiple trees trained on random subsets of the data and features. Let $\mathcal{D} = \{(\boldsymbol{X_{t}}, Y_{t})\}_{t = 1}^{N}$ denote the training dataset, where $\boldsymbol{X_{t}} = (Y_{t - 1}, Y_{t - 2}, \cdots, Y_{t - p}, Z_{1t}, Z_{2t}, \cdots, Z_{qt})^{\top}$ represents the predictor vector with $p$ lagged values and $q$ external variables. To enable a fair comparison with the ISOLR and TSOLR models, the external covariates are incorporated using two alternative seasonal representations, such as indicator variables and trigonometric terms, each evaluated separately. For each of the $K$ bootstrap samples $\mathcal{D}_b^{(k)}$, a classification tree $h_{k}(X)$ is fitted using a random subset of predictors at each split to reduce correlation among trees. The final Random Forest prediction is obtained through majority voting:
\[
\hat{y} = \arg\max_{j \in \mathcal{C}} \sum_{k = 1}^{K} \mathbb{I}\big(h_{k}(X) = j\big),
\]
where $\mathcal{C} = {0, 1, 2, \cdots, m}$ is the set of AQI categories and $\mathbb{I}(\cdot)$ is the indicator function. 

Random Forests are known for their robustness against overfitting, ability to model nonlinear and complex relationships and strong predictive performance even with high-dimensional categorical features. However, their ensemble nature reduces interpretability and increases computational cost, posing challenges for theoretical inference and causal analysis in time-dependent categorical data.

\subsection{Support Vector Machines (SVMs)}
Support Vector Machines (SVMs) are powerful supervised learning algorithms widely applied to classification problems (\cite{Awad2015}). For a categorical time series $\{Y_t\}$, the predictor vector at time $t$ may include both lagged responses as integers and exogenous covariates, i.e., $\boldsymbol{X_{t}} = (Y_{t-1}, \cdots, Y_{t-p}, Z_{1t}, \cdots, Z_{qt})^{\top}$. As in decision‐tree framework, we consider two alternative formulations of the covariates, one using seasonal indicator variables and the other employing trigonometric terms and fit the SVM models separately under each specification. Given training data $\mathcal{D} = \{(\boldsymbol{X_{t}}, Y_{t})\}_{t=1}^{N}$ with $Y_{t} \in \{-1,+1\}$, a standard SVM seeks a separating hyperplane $\boldsymbol{w}^\top \boldsymbol{X} + b = 0$ by solving
\begin{equation*}
    \min_{\boldsymbol{w}, b, \boldsymbol{\xi}} \frac{1}{2}\|\boldsymbol{w}\|^{2} + C\sum_{t = 1}^{N}\xi_{t}, \quad \text{subject to}\quad Y_{t}(\boldsymbol{w}^\top\boldsymbol{X_{t}} + b)\ge 1-\xi_{t},\ \xi_{t}\ge 0,
\end{equation*}
where $C>0$ regulates the balance between margin maximization and misclassification. For ordinal-valued $Y_t$ (e.g., AQI categories), ordinal SVMs generalize this framework by introducing a sequence of ordered parallel hyperplanes,
\begin{equation*}
    \boldsymbol{w}^\top \boldsymbol{X_{t}} - b_{1} = 0,\;\boldsymbol{w}^\top \boldsymbol{X_{t}} - b_{2} = 0,\;\ldots,\;\boldsymbol{w}^\top \boldsymbol{X_{t}} - b_{m - 1} = 0,
\end{equation*}
ensuring that predicted classes respect their natural ordering. By incorporating both lagged values and external variables into $\boldsymbol{X_{t}}$, ordinal SVMs can capture short-term temporal dynamics along with covariate-driven effects in categorical time series forecasting.

SVMs demonstrate strong generalization in high-dimensional feature spaces and are robust against overfitting. However, their performance depends critically on the choice of kernel function and regularization parameters. Moreover, computational complexity increases with sample size and SVMs do not inherently capture sequential dynamics beyond fixed lag structures, limiting their ability to model long-range temporal dependence.

\subsection{Recurrent Neural Networks (RNNs)}
Recurrent Neural Networks (RNNs) and their advanced variants, such as Long Short-Term Memory (LSTM) and Gated Recurrent Unit (GRU) architectures, are widely used for modeling temporal dependencies in sequential data (\cite{Lindemann2021}). Unlike traditional feedforward networks, RNNs incorporate feedback connections that allow information to persist across time steps. In this setting, the predictor sequence at time $t$ may include both lagged category indicators as integers and exogenous variables, i.e., $\boldsymbol{X_{t}} = (Y_{t-1}, \cdots, Y_{t-p}, Z_{1t}, \cdots, Z_{qt})^{\top}$. Given an input sequence $\{\boldsymbol{X_{t}}\}_{t = 1}^{N}$, a standard RNN updates its hidden state according to
\begin{equation*}
    \boldsymbol{h}_t = \sigma\left( W_h \boldsymbol{h}_{t-1} + W_x \boldsymbol{X_{t}} + \boldsymbol{b_{h}} \right),
\end{equation*}
where $\boldsymbol{h_{t}}$ denotes the hidden representation at time $t$, $\sigma(\cdot)$ is a nonlinear activation function and $W_{h}$, $W_{x}$, $\boldsymbol{b_{h}}$ are the model parameters. LSTMs enhance this framework by introducing gating mechanisms that regulate information flow and mitigate the vanishing gradient problem. Specifically, for each time step $t$, the LSTM computes
\[
\begin{aligned}
\boldsymbol{f_{t}} &= \sigma(W_{f} \boldsymbol{X_{t}} + U_{f} \boldsymbol{h_{t - 1}} + \boldsymbol{b_{f}}), \\
\boldsymbol{i_{t}} &= \sigma(W_{i} \boldsymbol{X_{t}} + U_{i} \boldsymbol{h_{t - 1}} + \boldsymbol{b_{i}}), \\
\tilde{\boldsymbol{c_{t}}} &= \tanh(W_{c} \boldsymbol{X_{t}} + U_{c} \boldsymbol{h_{t - 1}} + \boldsymbol{b_{c}}), \\
\boldsymbol{c_{t}} &= \boldsymbol{f_{t}} \odot \boldsymbol{c_{t - 1}} + \boldsymbol{i_{t}} \odot \tilde{\boldsymbol{c_{t}}}, \\
\boldsymbol{o_{t}} &= \sigma(W_{o} \boldsymbol{X_{t}} + U_{o} \boldsymbol{h_{t - 1}} + \boldsymbol{b_{o}}), \\
\boldsymbol{h_{t}} &= \boldsymbol{o_{t}} \odot \tanh(\boldsymbol{c_{t}}),
\end{aligned}
\]
where $\boldsymbol{f_{t}}$, $\boldsymbol{i_{t}}$ and $\boldsymbol{o_{t}}$ denote the forget, input and output gates respectively, $\odot$ represents element-wise multiplication and $\boldsymbol{c_{t}}$ is the memory cell state. Like decision tree and SVM, LSTM is also fitted using both types of covariates, namely, indicator variables and trigonometric terms as seasonal components.

For ordinal categorical outputs, a softmax output layer can be replaced or modified with an ordinal-aware loss, such as ordinal cross-entropy or cumulative link-based objectives, to preserve the intrinsic order among categories. Such architectures effectively capture nonlinear dynamics and long-range temporal dependencies, making them highly suitable for forecasting AQI categories over time. However, these models require extensive hyperparameter tuning to balance model complexity and generalization. Training can be computationally intensive due to sequential dependencies and like other deep learning methods, interpretability remains a challenge despite their strong predictive capability.

These machine learning methods offer flexible frameworks for modeling ordinal categorical time series and can be chosen based on data characteristics, interpretability requirements and computational considerations.

\section{Auto-Dependency Measures for Categorical Time Series}    \label{sec:autoassociation}
Autocorrelation function (ACF) is a standard tool for measuring serial dependence in numeric time series. ACF plots are also used rigorously to determine the suitable order of a model. However, it is not directly applicable to categorical data due to the absence of a natural arithmetic structure. To study dependence in categorical time series, especially when data are nominal or ordinal, several ACF-like measures have been proposed. 

Let $Y_{t} \in \{0, 1, \cdots, m\}$ be a categorical time series. Define the joint probability at lag $h$ as \( p_{ij}^{(h)} = P(Y_{t} = i, Y_{t - h} = j) \), the marginal probabilities as \( p_{i\cdot}^{(h)} = P(Y_{t - h} = i)  = \sum_{j = 0}^{m} p_{ij}^{(h)} \) and \( p_{\cdot j}^{(h)} = P(Y_{t - h} = j) = \sum_{i = 0}^{m} p_{ij}^{(h)} \). Under this setup, \cite{Weiss2008} summarized various auto-association measures, which are provided in Table \ref{tab:auto-association}. In the present analysis, these auto-association measures are computed not only to assess the strength of serial dependence in the categorical time series but also to guide the selection of an appropriate model order for subsequent time-series modeling.

\begin{table}[htpb]
    \centering
    \caption{Various auto-association measures for categorical time series}
    \begin{tabular}{|c|c|c|}
    \hline
        Measure & Definition & Range \\ \hline
        Cohen's $\kappa$ & $\kappa(h) = \frac{\sum\limits_{i=0}^{m} p_{ii}^{(h)} - \sum\limits_{i=0}^{m} p_{i\cdot}^{(h)} p_{i\cdot}^{(h)}} {1 - \sum\limits_{i=0}^{m} p_{i\cdot}^{(h)} p_{i\cdot}^{(h)}}$ & [$- \frac{\sum\limits_{i=0}^{m} p_{i\cdot}^{(h)} p_{i\cdot}^{(h)}}{1 - \sum\limits_{i=0}^{m} p_{i\cdot}^{(h)} p_{i\cdot}^{(h)}}$, 1] \\ \hline
        Cramer's $\nu$ & $\nu(h) = \sqrt{\frac{1}{nm} \sum_{i = 0}^{m} \sum_{j = 0}^{m} \frac{(N_{ij}^{(h)} - E_{ij})^2}{E_{ij}}}$ & [0, 1] \\ \hline
        Goodman-Krushkal's $\tau$ & $A_{\nu}^{(\tau)}(h) = \frac{\sum_{i, j = 0}^{m}\frac{p_{ij}^{(h)}{p_{ij}^{{(h)}}}}{p_{\cdot j}^{(h)}} - \sum_{i = 0}^{m}p_{i\cdot}^{(h)}p_{i\cdot}^{(h)}}{1 - \sum_{i = 0}^{m}p_{i\cdot}^{(h)}p_{i\cdot}^{(h)}}$ & [0, 1] \\ \hline
        Goodman-Krushkal's $\gamma$ & $\gamma = \frac{\text{no. of concordant pairs } - \text{ no. of discordant pairs}}{\text{no. of concordant pairs } + \text{ no. of discordant pairs}}$ & [$-$1, 1] \\ \hline
        Pearson Measure $X_{n}^{2}$ & $X_{n}^{2} = n \sum_{i=0}^{m} \sum_{j=0}^{m} \frac{(\hat{p}_{ij}^{(h)} - \hat{p}_{i\cdot}^{(h)} \hat{p}_{\cdot j}^{(h)})^2}{\hat{p}_{i\cdot}^{(h)} \hat{p}_{\cdot j}^{(h)}}$ & [0, $n(m - 1)$] \\ \hline
        Mutual Information & $A_{\nu}^{(u)}(h) = -\frac{\sum_{i, j = 0}^{m} p_{ij}^{(h)}\ln{\left(\frac{p_{ij}^{(h)}}{p_{i\cdot}^{(h)}\,p_{\cdot j}^{(h)}}\right)}}{\sum_{i = 0}^m p_{i\cdot}^{(h)}\ln{p_{i\cdot}^{(h)}}}$ & [0, 1] \\ \hline
    \end{tabular}
    \label{tab:auto-association}
\end{table}

\section{Simulation Study}      \label{sec:simulation}
In this section, we conduct extensive simulation studies to check the consistency and forecasting performance of the TSOLR model. The data are generated using both the ISOLR and TSOLR models, under a variety of parameter configurations, as detailed in the following sections. The estimates are obtained through Maximum Likelihood Estimation (MLE) as described in Section \ref{sec:mle}.

We consider three different sample sizes ($n$) of 500, 1000, 10000. Each simulation setup is replicated 1000 times to ensure stable and reliable estimates.
\subsection{Empirical Consistency}
To examine the empirical consistency of the TSOLR model (\ref{equ:logistic_fourier}), we generate samples of different sizes from model (\ref{equ:logistic_fourier_sim}) with $P_{1} = 100$, $P_{2} = 5$ and $P_{3} = 100$, which is a particular case of model (\ref{equ:logistic_fourier}).
\begin{align}   \label{equ:logistic_fourier_sim}
    \log \left( \frac{P(Y_{t} \leq j)}{P(Y_{t} > j)} \right) = \ 
    & \theta_{j} -  \Big[\sum_{i = 1}^{2} \Bigl\{ \beta_{1i} \cdot \sin \left( \frac{2 \pi t}{P_{i}} \right) + \beta_{2i} \cdot \cos \left( \frac{2\pi t}{P_{i}} \right) \Bigl\} + \notag \\ 
    & \beta_{13} \cdot |\sin \left( \frac{2 \pi t}{P_{3}} \right)| + \beta_{23} \cdot |\cos \left( \frac{2 \pi t}{P_{3}} \right)| + \gamma Y_{t - 1}\Big].
\end{align}

In the data-generating process, two different parameter settings are considered: three-category case with parameters ($\theta_{0}$, $\theta_{1}$, $\beta_{11}$, $\beta_{21}$, $\beta_{12}$, $\beta_{22}$, $\beta_{13}$, $\beta_{23}$, $\gamma$) = ($-$0.84, 0.84, 0, 2, 0, 0.7, 0, 0.2, 1.9) and four-category case with parameters ($\theta_{0}$, $\theta_{1}$, $\theta_{2}$, $\beta_{11}$, $\beta_{21}$, $\beta_{12}$, $\beta_{22}$, $\beta_{13}$, $\beta_{23}$, $\gamma$) = ($-$1.1, 0, 1.1, 0.6, $-$1.1, 1.4, $-$0.8, $-$0.1, $-$0.4, $-$0.3). The simulation results, summarized in Tables \ref{tab:sim_con_1} and \ref{tab:sim_con_2}, show that the standard errors of the estimated parameters decrease as sample size increases.  This trend provides strong evidence of the empirical consistency of the MLEs, thereby confirming the reliability of the TSOLR model across a range of data sizes and categorical configurations.
\begin{table}[htpb]
    \centering
    \caption{Average estimated values of parameters along with the standard errors in parenthesis for true parameter values ($\theta_{0}$, $\theta_{1}$, $\beta_{11}$, $\beta_{21}$, $\beta_{12}$, $\beta_{22}$, $\beta_{13}$, $\beta_{23}$, $\gamma$) = ($-$0.84, 0.84, 0, 2, 0, 0.7, 0, 0.2, 1.9) in three-category case}
    \begin{tabular}{|c|c|c|c|}
    \hline
        $n$          & 500            & 1000           & 10000          \\ \hline
        $\theta_{0}$ & $-$1.12 (0.56) & $-$0.97 (0.33) & $-$0.85 (0.10) \\ \hline
        $\theta_{1}$ & 0.60 (0.58)    & 0.74 (0.34)    & 0.83 (0.10)    \\ \hline
        $\beta_{11}$ & 0.00 (0.29)    & 0.00 (0.18)    & 0.00 (0.06)    \\ \hline
        $\beta_{21}$ & 2.27 (0.69)    & 2.12 (0.40)    & 2.01 (0.11)    \\ \hline
        $\beta_{12}$ & 0.00 (0.29)    & $-$0.01 (0.19) & 0.00 (0.06)    \\ \hline
        $\beta_{22}$ & 0.81 (0.38)    & 0.75 (0.23)    & 0.70 (0.07)    \\ \hline
        $\beta_{13}$ & $-$0.01 (0.21) & 0.00 (0.14)    & 0.00 (0.04)    \\ \hline
        $\beta_{23}$ & 0.21 (0.20)    & 0.21 (0.14)    & 0.20 (0.05)    \\ \hline
        $\gamma$     & 1.88 (0.21)    & 1.89 (0.15)    & 1.90 (0.04)    \\ \hline
    \end{tabular}
    \label{tab:sim_con_1}
\end{table}
\begin{table}[htpb]
    \centering
    \caption{Average estimated values of parameters along with the standard errors in parenthesis for true parameter values ($\theta_{0}$, $\theta_{1}$, $\theta_{2}$, $\beta_{11}$, $\beta_{21}$, $\beta_{12}$, $\beta_{22}$, $\beta_{13}$, $\beta_{23}$, $\gamma$) = ($-$1.1, 0, 1.1, 0.6, $-$1.1, 1.4, $-$0.8, $-$0.1, $-$0.4, $-$0.3) in four-category case}
    \begin{tabular}{|c|c|c|c|}
    \hline
        $n$          & 500             & 1000            & 10000          \\ \hline
        $\theta_{0}$ & $-$1.14 (0.16)  & $-$1.12 (0.12)  & $-$1.10 (0.04) \\ \hline
        $\theta_{1}$ & $-$0.02 (0.15)  & $-$0.01 (0.11)  & 0.00 (0.03)    \\ \hline
        $\theta_{2}$ & $-$0.02 (0.16)  & $-$0.01 (0.11)  & 0.00 (0.05)    \\ \hline
        $\beta_{11}$ & 0.61 (0.13)     & 0.61 (0.09)     & 0.60 (0.03)    \\ \hline
        $\beta_{21}$ & $-$1.13 (0.14)  & $-$1.11 (0.10)  & $-$1.10 (0.03) \\ \hline
        $\beta_{12}$ & 1.43 (0.15)     & 1.42 (0.10)     & 1.40 (0.03)    \\ \hline
        $\beta_{22}$ & $-$0.82 (0.14)  & $-$0.81 (0.09)  & $-$0.80 (0.03) \\ \hline
        $\beta_{13}$ & $-$0.10 (0.13)  & $-$0.10 (0.09)  & $-$0.10 (0.03) \\ \hline
        $\beta_{23}$ & $-$0.41 (0.12)  & $-$0.40 (0.09)  & $-$0.40 (0.03) \\ \hline
        $\gamma$     & $-$0.33 (0.08)  & $-$0.31 (0.06)  & $-$0.30 (0.02) \\ \hline
    \end{tabular}
    \label{tab:sim_con_2}
\end{table}
\subsection{Forecasting Performance}
To assess the forecasting performance of the TSOLR model, we compare the forecasting performance of TSOLR model with ISOLR model.

We generate samples of different sizes from the model (\ref{equ:logistic_indicator}), incorporating three types of seasonal indicators and a lagged response. Specifically, the seasonality structure includes a three-level categorical variable $S_{1t}$, representing seasons (e.g., Season 1, Season 2 and Season 3) and two binary indicators, $S_{2t}$ and $S_{3t}$, corresponding to festive period and type of day, respectively. As the Season variable has three levels, we create two dummy variables, $S_{1t}^{(1)}$ and $S_{2t}^{(2)}$ corresponding to Season 1 and Season 2 respectively, as follows.
\begin{equation*}
    S_{1t}^{(1)} = 
    \begin{cases}
        1 & \text{if } t \in [100k + 1, 100k + 30] \\
        0 & otherwise
    \end{cases}
\end{equation*}
and
\begin{equation*}
    S_{1t}^{(2)} = 
    \begin{cases}
        1 & \text{if } t \in [100k + 31, 100k + 70] \\
        0 & otherwise
    \end{cases}
\end{equation*}
for $k = 0, 1, 2, \cdots, \frac{n}{100} - 1$. Similarly $S_{2t}$ and $S_{3t}$ are defined as follows.
\begin{equation*}
    S_{2t} = 
    \begin{cases}
        1 & \text{if } t \in [5k + 4, 5k + 5] \\
        0 & otherwise
    \end{cases}
\end{equation*}
and
\begin{equation*}
    S_{3t} = 
    \begin{cases}
        1 & \text{if } t \in [100k + 81, 100k + 85] \\
        0 & otherwise
    \end{cases}
\end{equation*}
for $k = 0, 1, 2, \cdots, \frac{n}{100} - 1$. That means, first 30 days of each 100-day period (which is equivalent to 1 year period in this simulation) are assigned as Season 1, next 40 days (i.e., $31^{st}$ to $70^{th}$ day) are assigned as Season  2 and rest 30 days are assigned as Season 3. Similarly, $S_{2t}$ takes value 1 for 4th and 5th day for each 5-day period (which is equivalent to 1 week period in this simulation) and $S_{3t}$ takes value 1 for $81^{st}$ to $85^{th}$ day for each 100-day period. The data-generating model is thus specified as
\begin{equation}    \label{equ:logistic_indicator_sim}
    \log \left( \frac{P(Y_{t} \leq j)}{P(Y_{t} > j)} \right) = \theta_{j} - \left( \beta_{1} \cdot S_{1t}^{(1)} + \beta_{2} \cdot S_{1t}^{(2)} + \beta_{3} \cdot S_{2t} + \beta_{4} \cdot S_{3t} + \beta_{5} \cdot Y_{t - 1} \right).
\end{equation}

In the data-generating process, two different parameter settings are considered: three-category case with parameters ($\theta_{0}$, $\theta_{1}$, $\beta_{1}$, $\beta_{2}$, $\beta_{3}$, $\beta_{4}$, $\beta_{5}$) = ($-$0.8, 0.78, 0.2, $-$1.4, 0.3, 0.9, 1.1) and four-category case with parameters ($\theta_{0}$, $\theta_{1}$, $\theta_{2}$, $\beta_{1}$, $\beta_{2}$, $\beta_{3}$, $\beta_{4}$, $\beta_{5}$) = ($-$0.9, 0.02, 1.2, $-$0.4, $-$1.3, 0.5, 0.7, 1.2). A training set of size 80\% of the data is used to fit both of the models (\ref{equ:logistic_fourier_sim}) and (\ref{equ:logistic_indicator_sim}) and a test set of size 20\% of the data is used to make 1-step ahead forecast and compute the accuracy and weighted F1 score, weights being the fraction of different categories in the test set. The results are presented in Tables \ref{tab:sim_for_1} and \ref{tab:sim_for_2}. These results show that though the ISOLR model outperforms the TSOLR model, the differences are very small. Also the differences are decreasing with increment in sample size. This establishes the fact that TSOLR model performs well enough in forecasting ordinal categorical data with seasonal effects.
\begin{table}[htpb]
    \centering
    \caption{Average accuracy and weighted F1 score for forecasting with true parameter values ($\theta_{0}$, $\theta_{1}$, $\beta_{11}$, $\beta_{21}$, $\beta_{12}$, $\beta_{22}$, $\beta_{13}$, $\beta_{23}$, $\gamma$) = ($-$0.8, 0.78, 0.2, $-$1.4, 0.3, 0.9, 1.1) in three-category case}
    \begin{tabular}{|c|cc|cc|}
    \hline
    \multirow{2}{*}{$n$} & \multicolumn{2}{c|}{ISOLR model} & \multicolumn{2}{c|}{TSOLR model} \\ \cline{2-5}
                         & Accuracy & Weighted F1 & Accuracy & Weighted F1 \\ \hline
    500                  & 67.53\%  & 63.16\%     & 66.89\%  & 62.09\%     \\ \hline
    1000                 & 67.29\%  & 63.18\%     & 66.92\%  & 62.26\%     \\ \hline
    10000                & 67.50\%  & 63.83\%     & 67.20\%  & 63.00\%     \\ \hline
    \end{tabular}
    \label{tab:sim_for_1}
\end{table}
\begin{table}[htpb]
    \centering
    \caption{Average accuracy and weighted F1 score for forecasting with true parameter values ($\theta_{0}$, $\theta_{1}$, $\theta_{2}$, $\beta_{11}$, $\beta_{21}$, $\beta_{12}$, $\beta_{22}$, $\beta_{13}$, $\beta_{23}$, $\gamma$) = ($-$0.9, 0.02, 1.2, $-$0.4, $-$1.3, 0.5, 0.7, 1.2) in four-category case}
    \begin{tabular}{|c|cc|cc|}
    \hline
    \multirow{2}{*}{$n$} & \multicolumn{2}{c|}{ISOLR model} & \multicolumn{2}{c|}{TSOLR model} \\ \cline{2-5}
                         & Accuracy & Weighted F1 & Accuracy & Weighted F1 \\ \hline
    500                  & 79.07\%  & 72.61\%     & 78.89\%  & 72.45\%     \\ \hline
    1000                 & 79.05\%  & 72.46\%     & 78.96\%  & 72.35\%     \\ \hline
    10000                & 79.28\%  & 72.34\%     & 79.21\%  & 72.44\%     \\ \hline
    \end{tabular}
    \label{tab:sim_for_2}
\end{table}

\section{Data Analysis Results}       \label{sec:results}

\subsection{Order Selection through Auto-association Measures}
\begin{figure}[htpb]
    \centering
    \includegraphics[scale = 0.9]{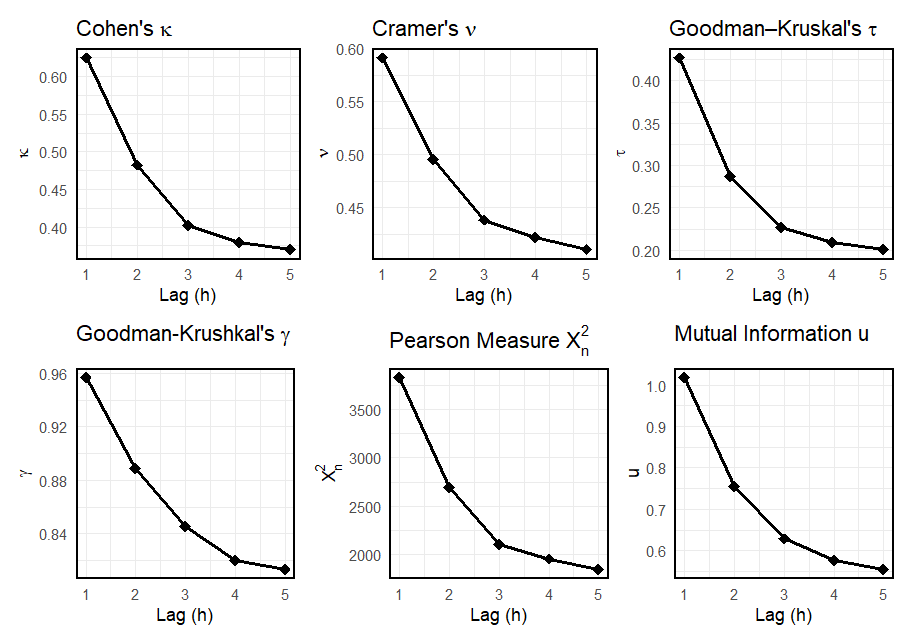}
    \caption{Auto-association measures of the categorical AQI time series computed over varying lag values}
    \label{fig:auto-association}
\end{figure}

Various lag-wise dependency measures, as depicted in Table \ref{tab:auto-association}, are computed to assess autocorrelation and potential model order. The plot in Figure \ref{fig:auto-association} illustrates a clear decreasing trend in auto-association, indicating diminishing temporal dependence as the lag increases. These insights guided the choice of first-order models for further analysis.

\subsection{Forecasting Comparison}
To assess the effectiveness of various forecasting approaches for categorical AQI prediction, the data from 2019 to 2023 were used as the training set, while the data of 2024 served as the test set. We evaluated the models discussed in Section \ref{sec:logistic} and \ref{sec:methods}. Each model was trained to predict the AQI category of next day based on seasonal information and current AQI state. The ISOLR model includes discrete seasonal indicators in the linear predictor as given in model (\ref{equ:logistic_indicator}). This model is highly interpretable and captures categorical effects from major seasonal patterns. The TSOLR model uses sinusoidal terms to model seasonality more smoothly as in model (\ref{equ:logistic_fourier}). This version captures both weekly and annual periodic components, including higher-frequency events such as Diwali.

To complement these models, classical time series models like MTD and PAR models were also implemented. These models does not take the external covariates like seasonal terms into account. The forecasting using these models were based only on the lag-1 values.

Apart from these models, we also implemented ML and DL models like Random Forest, Support Vector Machines and LSTM. These models were trained using both the indicator variables and trigonometric variables seperately along with the lag-1 AQI categories. Although not designed for ordinal outcomes, they provide flexible non-linear classification capabilities and serve as useful benchmarks.

\begin{figure}[htpb]
    \centering
    \includegraphics[scale = 0.8]{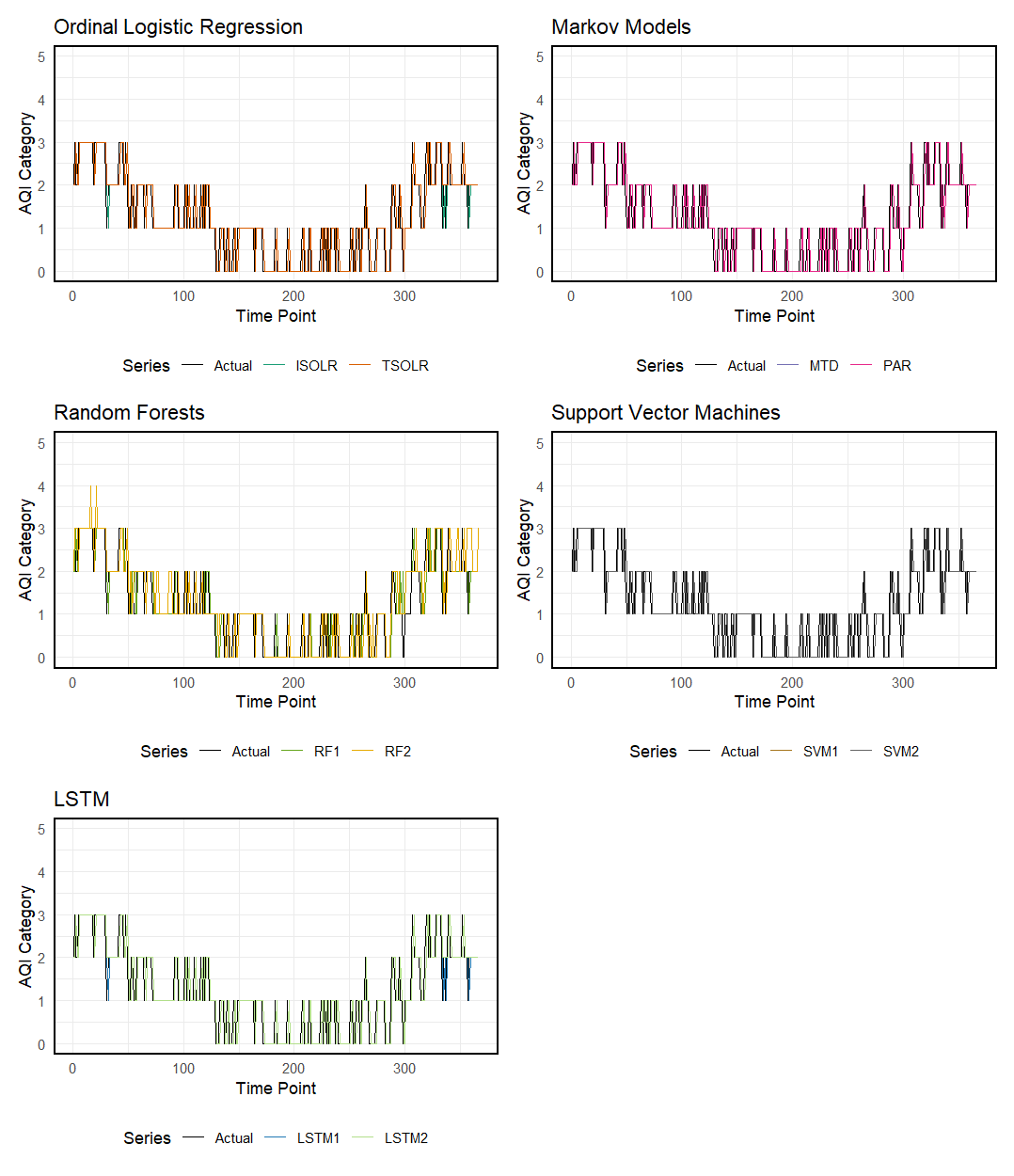}
    \caption{Predicted AQI categories for various models along with the true categories; 1 and 2 represent the indicator based and Fourier series based model respectively}
    \label{fig:forecasts}
\end{figure}

The forecasting results for Kolkata, as presented in Table \ref{tab:data_analysis_forecast}, indicate that the TSOLR model achieves the highest overall performance, with an accuracy of 74.04\% and a weighted F1 score of 74.07\%. Among all models compared, including traditional time series models (MTD(1), PAR(1)), machine learning methods (Random Forest, Support Vector Machine) and deep learning (LSTM), the Fourier series based logistic and SVM models consistently outperform their indicator based counterparts. Notably, while Random Forest performs well with indicator features, its performance drops significantly when Fourier features are used. Figure \ref{fig:forecasts} shows the actual AQI categories for the test set along with predicted categories for different models. For the last three panels of Figure \ref{fig:forecasts}, 1 and 2 represent indicator based model and trigonometric series based model respectively. These highlight the effectiveness of Fourier series based seasonality encoding in capturing underlying temporal patterns, particularly when combined with models like logistic regression or SVM, making them suitable choices for categorical time series forecasting in this context.

\begin{table}[htpb]
\caption{Forecasting results for Kolkata data analysis}
\label{tab:data_analysis_forecast}
\centering
  \begin{tabular}{|c|c|c|}
    \cline{1-3}
    Models                                        & Accuracy         & Weighted F1 \\ \hline
    ISOLR                    & 72.95\%          & 72.96\% \\
    \textbf{TSOLR}      & \textbf{74.04\%} & \textbf{74.07\%} \\
    MTD(1)                                        & 72.68\%          & 72.68\% \\    
    PAR(1)                                        & 72.68\%          & 72.68\% \\
    Random Forest (Indicator based)               & 71.31\%          & 71.38\% \\
    Random Forest (Fourier series based)          & 65.85\%          & 66.02\% \\
    Support Vector Machine (Indicator based)      & 72.68\%          & 72.68\% \\
    Support Vector Machine (Fourier series based) & 73.50\%          & 73.51\% \\
    LSTM (Indicator based)                        & 72.88\%          & 72.89\% \\ 
    LSTM (Fourier series based)                   & 73.70\%          & 73.71\% \\ \hline
  \end{tabular}
\end{table}

\section{Discussion}    \label{sec:discussion}
This study investigates the effectiveness of ordinal logistic regression models for short-term forecasting of categorical Air Quality Index (AQI) levels using daily data. The TBATS model (\cite{Livera2011}) employs trigonometric Fourier series expansions to capture multiple and potentially complex seasonal patterns in continuous time series data. Motivated by this framework, we adapt the underlying idea to the ordinal setting by incorporating trigonometric seasonal components within an ordinal logistic regression model (TSOLR), allowing for smoothly varying seasonal structures in categorical time series. In parallel, we also develop an alternative specification that replaces the Fourier terms with seasonal indicator variables (ISOLR), enabling the model to capture abrupt or discrete seasonal shifts. These two complementary formulations provide flexible approaches for representing both gradual and sudden seasonal variations in ordinal AQI data. By incorporating seasonal components through both indicator variables and Fourier series terms, the proposed models provide a flexible framework to capture the temporal dynamics and periodicity inherent in AQI data.

The ISOLR model allows intuitive interpretation of the effects of categorical seasonal variables such as various seasons, weekends and festival periods like Diwali. In contrast, the TSOLR model captures continuous seasonal cycles and higher-frequency fluctuations, allowing for smoother estimation of both annual and weekly variations, including sharp spikes during festivals. 

We have conducted comprehensive simulation experiments that have demonstrated the empirical consistency of the model parameters using the MLE method. Furthermore, though the ISOLR model outperforms the TSOLR model, the differences are very small. Also the differences are decreasing with increment in sample size. That shows the TSOLR model performs reasonably well for the purpose of forecasting of ordinal categorical time series with multiple seasonalities.

The categorical time series analysis of AQI data for Kolkata demonstrates that both approaches are effective for 1-day ahead forecasting, but the Fourier-based model slightly outperforms the indicator-based one in terms of overall prediction accuracy across multiple forecast horizons. This advantage is likely due to its ability to capture subtle and complex seasonal structures, especially higher frequency events. Comparisons with machine learning models such as Random Forests and Support Vector Machines (SVM) show that while these methods offer competitive performance, the proposed ordinal logistic regression models provide the added benefit of interpretability, formal probabilistic structure and ease of incorporating prior knowledge through structured seasonal terms. The use of auto-association measures for lag selection further strengthened the modelling process by identifying meaningful dependencies across time.

This analysis opens several avenues for further research. First, the proposed models can be extended to incorporate spatial dependencies, for instance, by jointly modeling AQI data from multiple monitoring stations using spatially correlated random effects. Second, the inclusion of additional time-varying exogenous variables, such as wind speed, humidity and particulate matter ($PM_{2.5}$ and/or $PM_{10}$) concentrations, could enhance predictive power. Third, relaxing the proportional odds assumption using partial or non-proportional odds models may improve the model’s flexibility when predictor effects differ across AQI thresholds.

Additionally, Bayesian estimation of the proposed models with hierarchical priors on seasonal effects or Fourier coefficients could provide robust inference in the presence of noisy or sparse data. Finally, integrating the proposed framework into a real-time decision support system for urban environmental policy would offer practical utility in pollution control and public health planning.

\section*{Acknowledgement}
The author would like to thank OpenAI as ChatGPT GPT-4 by OpenAI (\url{https://chatgpt.com}) was used and prompts like ``Please polish the paragraph for clarity and grammatical mistakes" were provided to assist in language editing, grammar correction and improving the clarity of expression in the manuscript. Any kind of generative AI or LLM is not used to generate any intellectual and analytical content like original ideas, analyze data or draw conclusions. The authors would also like to thank the anonymous reviewers for their valuable comments and constructive suggestions, which have greatly improved the quality and clarity of this paper.

\section*{Funding Statement}
No external funding was received for this study.

\section*{Conflict of Interest Statement}
Authors declare that there is no conflict of interest related to this study.

\section*{Data Availability Statement}
The data that support the findings of this study are available in GitHub at \url{https://github.com/ganirban004/aqi-logistic-paper.git}. These data were derived from the website of Central Pollution Control Board, India (\url{https://airquality.cpcb.gov.in/ccr/#/caaqm-dashboard-all/caaqm-landing/aqi-repository}), which is publicly available.

\printbibliography

\end{document}